\documentclass[epj,draft]{svjour}
\usepackage[]{psfig}

\begin{document}
\title{Mach-Zehnder fiber interferometer test of the anisotropy of the speed of light}

\author{Victor de Haan}

\institute{BonPhysics B.V., Laan van Heemstede 38, 3297 AJ Puttershoek, The Netherlands, \email{victor@bonphysics.nl}}

%
\date{Version: April 2009}
\abstract{Two optical fiber Mach-Zehnder interferometers were constructed in an environment with a temperature stabilization of better than 1~mK per day. One interferometer with a length of 12~m optical fiber in each arm with the main direction of the arms perpendicular to each other. Another with a length of 2~m optical fiber in each arm where the main direction of the arms are parallel as a control. In each arm 1~m of fiber was wound around a ring made of piezo material enabling the control of the length of the arms by means of a voltage. The influence of the temperature on the optical phase difference between the interferometer arms was measured. It is attributed to the temperature change induced variation of the interaction region of the optical fiber couplers. Further, the influence of rotation of the interferometers at the Earth surface on the observed phase differences was determined. For one interferometer (with the long and perpendicular arms) it was found that the phase difference depends on the azimuth of the interferometer. For the other one (with the short and parallel arms) no relevant dependence on the azimuth has been measured.   
} 

\maketitle

\section{Introduction}

Since at the end of the 19th century methods became accurate enough to measure the speed of light, experiments were devised to measure the anisotropy of the speed of light at the Earth surface. This was sought to be done by so-called first order experiments, where the effect depends in first order on the ratio $v/c$, where $v$ is the velocity of the observer with respect to a preferred rest frame and $c$ is the speed of light in this frame. When Fresnel introduced his famous Fresnel drag coefficient it was believed that all possible first order effects were compensated by an ether drag. Then Maxwell~\cite{Maxwell1880} came along with the notion of second order experiments, where the effect depends in second order on the same ratio. Although Maxwell thought at that time it would be beyond any means of experimental method to measure a second order effect, one year later in 1881 Michelson~\cite{Michelson1881} devised an apparatus that should be able to measure the change of the velocity very accurately. The apparatus is now known as a Michelson-Morley interferometer. After some comments on the experiment by Lorentz in 1886~\cite{Lorentz1886} Michelson and Morley~\cite{Michelson1887} increased the sensitivity of the apparatus with almost a factor of ten overcoming the accuracy objections of Lorentz. The accuracy of the apparatus was further increased with a factor of 6 by Morley and Miller~\cite{Morley1905} and by Miller in a series of experiments between 1905 and 1930~\cite{Miller1922,Miller1926,Miller1930,Miller1933}. In all these experiments the sought for magnitude of the effect was never observed. This is satisfactorily explained by the Lorentz-Fitzgerald contraction~\cite{Lorentz1895} or by Einsteins theory of relativity~\cite{Einstein1905}. However, Miller in his elaborate series of experiments, always claimed that he measured a small second order effect and also a first order effect. The second order effects he measured were quite small with respect to the sought for effect, but larger than the experimental error. These second order effects were analysed by him by combining measurements at different epochs. Combining the results from these epochs and assuming the Sun moves relative to the preferred rest frame he was able to find a preferred direction in space and a velocity. The first order effect he measured depended very much on the detailed experimental settings and were not analysed to find an anisotropy. 

In February 1927 a conference on the experiment and theoretical background was held at the Mount Wilson Observatory~\cite{Conference1927}. This conference did not succeed in finding a flaw in either experiment or theory, leaving the discrepancy in tact. In 1955 Shankland, a former pupil of Miller, re-analysed Millers data~\cite{Shankland1955} and concluded that the second order effects do exist and remarks that they \textit{remain essential constant in phase and amplitude through periods of several hours and are then associated with a constant temperature pattern in the observation hut}. Assuming that during several hours the second order effect should change considerably, he then concludes that there is no second order effect and contributes any other changes to temperature effects. However, it was already shown by Miller~\cite{Miller1933} that the changes during several hours could be very small depending on the sidereal time and the epoch. Hence, the conclusion of Shankland is unsupported and the discrepancy between Millers results and theoretical expectations remains. 

At the end of the last century some new interests in the theory of the interferometric method to determine the anisotropy (or its absence) of the speed of light at the Earth surface emerged. M\'{u}nera~\cite{Munera1998} showed that the interpretation of the amplitude and phase of the second order effect should be done for each rotation of the interferometer separately, not by averaging on forehand. Further, following Hicks~\cite{Hicks1902} and Righi~\cite{Righi1919} De Miranda Filho describes possible first order effects in the Michelson-Morley interferometer~\cite{Filho2002}. This is quite a brave assumption as it is always believed that first order effects could not be detected with light interference measurements. This was also derived by Lorentz~\cite{Lorentz1886}. However, in his derivation he did not explicitly take into account the Doppler effect, as has been ignored by many of his contemporary researchers. Hicks~\cite{Hicks1902} and Righi~\cite{Righi1919} took the Doppler effect into account, finding a first order effect. In the above mentioned conference Lorentz acknowledges that the discrepancy between his derivation and that of Hicks or Righi should be resolved and he promises to do so. Unfortunately one year later he dies.

In view of this discrepancy several researchers try to find experimental evidence of first or second order effects with Michelson-Morley interferometer type instruments. This has been done by, for instance, Piccard~\cite{Piccard1926,Piccard1928}, Illingworth~\cite{Illingworth1927} and Joos~\cite{Joos1930}. All these authors report the absence of the sought for effect. However, according to M\'{u}nera these experiments all have results comparable with those of Miller. Hence, experimental evidence is not conclusive whether or not some first or second order effect exists. 

Recently M\'{u}nera~\cite{Munera2006} reported an experiment claiming to see second order effects. He used a Michelson-Morley interferometer being stationary in the laboratory frame. The rotation of the Earth was used to change the direction of the velocity of the apparatus with respect to the preferred frame. This idea was followed by Cahill~\cite{Cahill2008} using a fiber optical Mach-Zehnder interferometer. In these experiments the influence of the temperature on the signal was acknowledged. M\'{u}nera corrects his data for it and Cahill claims that the temperature can not influence the signal significantly. The present paper reports on an experiment where both the influence of temperature on a fiber Mach-Zehnder interferometer and the phase differences changes due to rotation around a vertical axis were measured, to determine the possible origin of the effect measured by Cahill. 

\section{Outline of the method}

\begin{figure}[tbp]
\begin{picture}(250,250)
\put(0,0){\psfig{figure=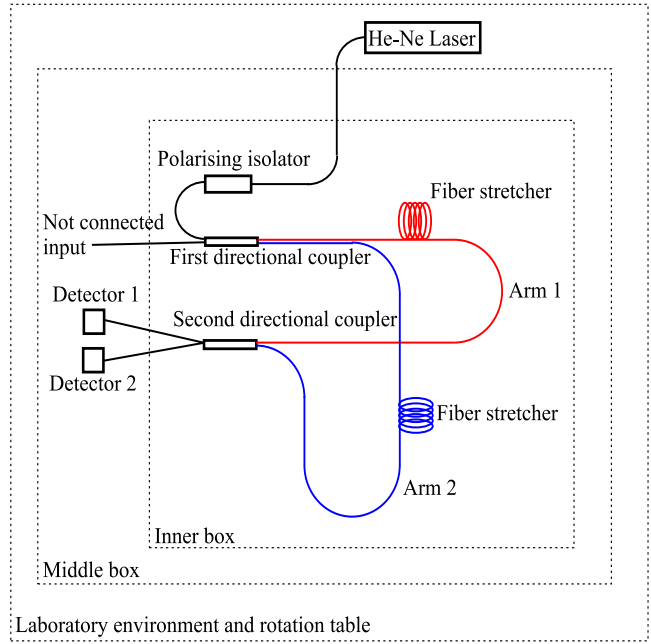,height=85mm,width=85mm}}
\end{picture}
\caption{\label{fig1} Configuration of the Mach-Zehnder interferometer.}
\end{figure}

With a fiber Mach-Zehnder interferometer (see figure~\ref{fig1}) the light injected into a fiber by means of a laser is split into two equal and in phase parts by a 2x2 directional coupler~\cite{BookDirCoupl}. These two light beams travel trough two (perpendicular) fiber arms of the interferometer and are rejoined by a second 2x2 directional coupler. Depending on their mutual phase, the light beams interfere constructively for one output fiber and destructively for the other one or vice versa. The two outputs of the second 2x2 directional coupler are fed into two detectors where their intensities $I_1$ and $I_2$ are measured. 
The sum of the intensities is proportional to the laser output power. It would be equal to the laser output power if all losses would be negligible. However, the coupling of the laser light into the fiber is not perfect, the absorption of the fiber is not nil and the coupling of the fibers are not perfect either. The difference of the two intensities relative to their sum is called the visibility, $V$. For an ideal interferometer the visibility would change between -1 and +1, depending on the phase difference, $\Delta \phi$ of the light beams in the second directional coupler according to
\begin{equation} \label{eq1}
 V = \frac{I_1-I_2}{I_1+I_2}=\cos{\Delta \phi} .
\end{equation}
The phase difference is determined by the optical path of the light, while traveling along the fibers from the first directional coupler to the second one. This optical path depends on the length of the arms and the wavelength of the light moving through the fibers of the arms. The wavelength, $\lambda$  of the light depends on the speed of light in the fibers and hence on the refractive index, $n$ of the fibers and (possibly) on the velocity and motion direction, $\vec{v}$ of the Earth relative to a preferred rest frame. Hence,
\begin{equation} \label{eq2}
\Delta \phi =\oint_{1} \frac{2\pi}{\lambda(n,\vec{v})}dl -  \oint_{2} \frac{2\pi}{\lambda(n,\vec{v})}dl ,
\end{equation}
where $\oint_{i}$ denotes the line integral along arm $i$ from the first directional coupler to the second one. The determination of the change in phase difference due to the motion of the Earth is the objective of this experiment. Therefore the whole interferometer is put on a rotation table to be able to change the azimuth of the interferometer. In this way the motion direction of the Earth relative to the direction of the arms changes during rotation. When tracking the phase difference changes during the rotation of the interferometer, one can get the magnitude of and the direction in which the projection of the Earth velocity with respect to the preferred rest frame is maximum. These values, relative to the local North, change with time of year and the sidereal day. If there is any effect due to the rotation of the Earth this should give a change in this signal depending on sidereal time (for a detailed discussion on the dependence on this effect see for instance~\cite{Munera1998}). Of course for a perfect interferometer it is not needed to rotate it as the Earth rotates around its axis every 24 hours. However, for a real-life fiber interferometer it is very difficult to get a stable signal for more than several hours, worse for days and virtual impossible for a year.

The polarisation properties of the light beams and the non-ideal directional couplers changes equation~(\ref{eq1}) into~\cite{BookFOS}:
\begin{equation} \label{eq1b}
V = \hat{V}\cos\{\Delta \phi + \phi \}+V_o , 
\end{equation}
where $\hat{V}$, $\phi$ and $V_o$ are determined by the exact properties of the directional couplers, the fibers used, beam polarisation and wavelength of the light. One of the problems of this equation is that the visibility is determined by the cosine of the phase difference between the light beams. There is an ambiguity considering the sign and exact value of $\Delta \phi$ when  only the visibility is determined. To overcome this problem an optical phase shifter is inserted in the arms of the interferometer, introducing an extra phase shift. This phase shifter is just an optical fiber wrapped around a thin walled cylinder made of piezo material. When applying a voltage over the wall of the cylinder the piezo material is either contracted or expanded, reducing or extending the length of the fiber wrapped around it. When the fiber stretches the optical path length changes accordingly and the phase of one light beam with respect to the other is changed. This added phase shift is controlled in such away that the visibility remains zero and hence the argument of the cosine remains $\pi/2$. The control signal can be calibrated and translated into an effective phase difference of the interferometer. In this way, also the sensitivity of the visibility to the polarisation direction of the light is greatly reduced, because when the visibility is reduced to zero $\hat{V}$ only determines the resulting accuracy of the control signal and not the control signal itself~\cite{BookFOS}.

\section{Influences on the phase difference} \label{secinf}

To determine the influences on the phase difference we simplify equation~(\ref{eq2}) by assuming the refractive index of the fiber, and the wavelength of the light in the fiber does not depend on the position in the fiber, then:
\begin{equation} \label{eq3}
\Delta \phi = \frac{2\pi n L_1}{\lambda_0} -  \frac{2\pi n L_2}{\lambda_0} , 
\end {equation}
where $n$ is the refractive index of the fibers used, $L_i$ is the length of arm $i$ and $\lambda_0$ the wavelength of the He-Ne laser in vacuum (633 nm).

Except for the sought for effect, several other factors can influence the phase difference of the light beams. For instance a change in the Earth magnetic field, in the relative humidity of the air or in the atmospheric pressure theoretically all can influence the phase difference. These influences have relatively small gradients and hence affect both arms in the same way. Hence, only the length difference between the arms must be taken into account. The experiment has been constructed with equal length fibers in the arms and hence the length difference, although not accurately measured, was much less than 1 cm. As the Earth magnetic field is very small and the fibers used are insensitive to magnetic fields the influence of the Earth magnetic field is neglected. Further, as the fibers are made of glass the influence of the relative humidity of air can also be neglected. 

According to~\cite{Giallorenzi1982} the influence due to a change in the atmospheric pressure can be calculated from the compressibility of the fiber core and cladding material. The pressure sensitivity of the optical phase in a fiber is defined as $\Delta \alpha/(\alpha \Delta P)$, where $\Delta \alpha$ is the change in the phase $\alpha$ due to a pressure change $\Delta P$. For fiber optic materials this pressure sensitivity depends on the core diameter and cladding diameter and has a value between $-10^{-6}$ and $-10^{-5}$ per bar. The phase change for a pressure change of 10 mbar and a difference in length of the two arms of 1 cm results for a wavelength of 633 nm in a change of phase between 1.5 mrad and 15 mrad. Although not quite negliable, it can be taken into account by measuring the atmospheric pressure and assuring that during the measurements the pressure change is well below 10 mbar.

Another important factor is the frequency, $\nu$ of the He-Ne laser. According to equation~(\ref{eq3}) the phase difference is determined by the wavelength of the light, generated by the laser. Although the frequency of the He-Ne laser is quite stable it can have mode changes due to changes in the length of the cavity introducing a frequency change. A standard laboratory He-Ne laser has a longitudinal mode sweep between 0.3 and 1.1 GHz. A stabilized laser reduces this mode sweep to 10 MHz or below. The influence due a change in frequency can be determined from equation~(\ref{eq3}) 
\begin{equation} \label{eq4a}
\frac{\partial \Delta \phi}{\partial \nu}  =  \frac{2\pi n \left(L_1-L_2\right)}{c}\left(1 + \frac{\partial n}{\partial \lambda_0}\frac{\lambda_0 }{n}\right) ,
\end {equation}
where it was used that $\nu=c/\lambda_0$ and $c$ is velocity of light in vacuum. The first term is due to the change of the wavelength of the light and the second term is due to the dispersion of the refractive index of the fiber material. For fuse silica at 300~K and a wavelength of 633~nm this dispersion is $\partial n/\partial \lambda_0 = -0.030 \ \mu $m$^{-1}$~\cite{Dispersion} so that it can be neglected with respect to the first term. Hence, for a length difference of 1 cm the phase change is approximately 0.305~rad/GHz.
 
The most important influence however is that of ambient temperature, $T$. Both the refractive index and the length of the fiber arms are dependent on the temperature. Then, from equation~(\ref{eq2}) for a homogeneous temperature change it follows that
\begin{equation} \label{eq4b}
\frac{\partial \Delta \phi}{\partial T}  =  \frac{2\pi n \left(L_1-L_2\right)}{\lambda_0} \left(  \alpha_n + \alpha_L\right) ,
\end {equation}
where $\alpha_n =\partial n /(n \partial T)$ and $\alpha_L=\partial L / (L\partial T)$. The first term represents the change in refractive index due to a temperature change and the second term represents the change in length of the fiber arms. For fused silica fibers as used here: $n=1.457$, $\alpha_n = 8.6 \times 10^{-6}$~1/K and $\alpha_L = 5.5 \times 10^{-7}$~1/K. Hence, for a length difference of 1~cm the phase change is approximately 1.3~rad/K. This gives an indication about the needed temperature stability of the interferometer. Again, when the temperature gradient is small both arms are affected in the same way and only the length difference between the arms must be taken into account. However, when the temperature gradient becomes too large this approximation does not hold, and the total length of the arms must be taken into account. Then, the phase change can easily grow to hundreds of radians per degree Celsius as instead of the length difference, the total length of the arms have to be taken into account. Hence, extreme care must be taken to avoid temperature gradients.

Another important temperature effect is that the properties of the directional coupler itself depend on the temperature via the coupling length, $l$ and the refractive index. These properties influence $\hat{V}$, $\phi$ and $V_0$. If the polarisation influence is neglected and only the coupling length of the directional couplers and a different absorption in both arms are taken into account, it is possible to determine equation~(\ref{eq1b}) by using the transfer matrix appraoch~\cite{BookDirCoupl}. Let $i_1$ and $i_2$ be the amplitudes of the light waves at the two inputs of the directinal coupler and $o_1$ and $o_2$ the same for the outputs, then for couplers with identical waveguides:
\begin{equation} \label{eq4c}
\left( \begin{array}{c} o_1  \\ o_2 \end{array} \right)=
\left[ \begin{array}{cc} \cos \kappa l_i & -i \sin \kappa l_i \\ -i\sin \kappa l_i & \cos \kappa l_i  \end{array} \right] 
\left( \begin{array}{c} i_1  \\ i_2 \end{array} \right) ,
\end{equation}
where $\kappa=2\pi n / \lambda_0$ and $l_i$ is the coupling length of either the first coupler or the second one. The 2$\times 2$~matrix is refered to as {\it transfer matrix} and describes the propagation of the beam through the coupler~\cite{BookDirCoupl}. The propafation of the beam through the arms of the interferometer can be described by a simple multiplication of the amplitude at the beginning of the arm by the transmission and a phase factor describing the optical path length. This can be written in another transfer matrix, that is diagonal as there is no interaction between the beams when travelling through the arms. The elements of the diagonal matrix can be written as $\tau_i \exp(-i \phi_i)$, where $\tau_i$ is the root of the transmission of the fiber (as we are dealing with amplitudes, not with intensities) and $\phi_i$ is optical path length of arm $i$. The complete transfer from the inputs of the first coupler (where $i_1=1$ and $i_2=0$) is given by
\begin{equation} \label{eq4d}
 \left( \begin{array}{c} o_1  \\ o_2 \end{array} \right)= 
 \left[ \begin{array}{cc} \cos \kappa l_2 & -i \sin \kappa l_2 \\ -i\sin \kappa l_2 & \cos \kappa l_2  \end{array} \right] \times
\end{equation}
 \[ \times
 \left[ \begin{array}{cc} \tau_1 e^{-i \phi_i} & 0 \\ 0 & \tau_2 e^{-i \phi_2}  \end{array} \right]
 \left[ \begin{array}{cc} \cos \kappa l_1 & -i \sin \kappa l_1 \\ -i\sin \kappa l_1 & \cos \kappa l_1  \end{array} \right]
 \left( \begin{array}{c} 1  \\ 0 \end{array} \right) .
\]
Taking $I_i$=$\left|o_i\right|^2$ and inserting them in equation~(\ref{eq1}), yields equation~(\ref{eq1b}) with
\begin{equation} \label{eq4e}
 \hat{V} = \frac{\sin{2\kappa l_1} \sin{2\kappa l_2}}{\eta^{-1}\sin^2{\kappa l_1}+\eta \cos^2{\kappa l_1}}
\end{equation}
and
\[
 V_o = \cos{2\kappa l_2}\frac{\eta^{-1}\sin^2{\kappa l_1}-\eta \cos^2{\kappa l_1}}{\eta^{-1}\sin^2{\kappa l_1}+\eta \cos^2{\kappa l_1}}
\]
where $\eta= \tau_1/\tau_2$, $\Delta \phi=\phi_1-\phi_2$ and $\phi=0$. Interestingly, if one of the couplers is an ideal 3dB directional coupler, with $2\kappa l_i=(m_i+1/2)\pi$, then $V_o = 0$. Here $m_i$ is an integer determined by the design of the directional coupler. However, for $\hat{V}$ to be $\pm 1$, both couplers should be ideal and the transmission of the arms should be equal. If the transmissions are not the same the amplitude of the visibilty change is reduced to $2/(\eta+\eta^{-1})$.
For non-ideal couplers a small offset for the visibilty can occur and a diminished amplitude. Let us assume that $2\kappa l_i=(m_i+1/2)\pi(1+\epsilon_i)$, where $\epsilon_i$ represents the influence of a change in temperature, frequency or another parameter. Then upto first order in $\epsilon_i$
\begin{equation} \label{eq4f}
 \hat{V} = \frac{2\cos{m_2\pi}}{\eta+\eta^{-1}}\left(\cos{m_1\pi}+\frac{\eta-\eta^{-1}}{\eta+\eta^{-1}}\pi(m_1+1/2)\epsilon_1\right)
\end{equation}
and
\[
 V_o =\frac{\eta-\eta^{-1}}{\eta+\eta^{-1}}\cos(m_2\pi)\pi(m_2+1/2)\epsilon_2 .
\]
From this equation is is clear that first order small changes can only influence the visibility via a change in transmission of one of the arms or when the transmission of the arms are not equal.

In the measurements a phase shifter is used, always adjusting the phase so that $\Delta \phi \approx \pi/2$. Small changes in $V_o$ and $\hat{V}$ due to temperature changes can occur, resulting in a shift of $\Delta \phi$. Hence,
\begin{equation}
\frac{\partial \Delta \phi}{\partial T} \approx  -\frac{\partial }{\partial T}
\left\{ \frac{V_0}{\hat{V}} \right\} ,
\end{equation}
using equation~(\ref{eq4f}), this becomes
\begin{equation}
\frac{\partial \Delta \phi}{\partial T} \approx -\frac{\pi(m_2+1/2)}{2\cos{m_1\pi}}\left\{ \frac{\epsilon_2}{1+\eta^{-2}}\frac{\partial \eta }{\partial T}  + (\eta-\eta^{-1})\frac{\partial  \epsilon_2}{\partial T} \right\} .
\end{equation}
The first term to the right describes the temperature influence via transmission change of the fibers. The transmission can change, but not more than a few precent so that this term can be neglected. The second term describers the temperature inflence via the change of either the length of the coupler or the refractive index. This term can be reduced by using $\partial  \epsilon_2/\partial T = \alpha_n+\alpha_L$ and $\pi(m_2+1/2)=2l_2\kappa = 4nl_2\pi/\lambda_0$ to
\begin{equation}
\frac{\partial \Delta \phi}{\partial T} \approx \pm \frac{2\pi nl_2 }{\lambda_0}(\eta-\eta^{-1})(\alpha_n+\alpha_L) ,
\end{equation}
where the $+$ holds when $m_2$ is odd and the $-$ for $m_2$ is even. The directional coupler is protected by a rigidly connected metal shield. The expansion coefficent of metal is larger than that of fiber glass increasing the temperature sensitivity. For aluminium for instance $\alpha_L = 23\times 10^{-6}$~1/K. The coupling length of the used couplers is approxiamtely 13~mm. When the relative transmission of one arm would be half of that of the other one ($\eta=\sqrt{2}$) the change in $\Delta \phi$ due to a temperature change is 42~mrad. Further, it should be noted that for the derivation of this formula it was assumed that the couplers were almost ideal and that polarisation effects do not play a role. Also it was assumed that the transfer matrix has the ideal form as given above. In reality these assumptions are not realized and the temperature influence might be much larger. It shows however, that altough at first sight the temperature has no or limited influence on the working of a directional coupler, in combination with a variation in another parameter, in this case the transmission difference of the arms, it can have an important influence.

From the above considerations it is clear that the effects of several influences on the phase difference can be kept under control as long as the gradient of these influence is so small as that they affect both arms of the interferometer in the same way. Then, only the length difference of the arms has to be taken into account reducing all influences to acceptable and controllable limits. See also table~\ref{table1} for a summary of the magnitude of the influences.

Influences that operate on one arm of the fiber alone could increase the influence to much higher values and must be avoided as much as possible. One, not so easy controlled, influence is stress on the fibers. Changes in stress can occur due to rotation of the whole set up, which might not be completely horizontal. Or by a change in temperature the support of the fibers might change its size and introduces stress. All these factors should be reduced as much as possible. The exact magnitude of the effect of stress on the fibers is hard to predict. The best way to determine them is by direct measurement. 

\begin{table}[bth]
\begin{tabular}{|l|l|l|} \hline
Parameter                       & Range                    & Influence   \\ 
                                &                          & mrad        \\ \hline 
Atmospheric Pressure            & 10 mbar                  & 1.5 - 15    \\  \hline
Ambient Temperature             & 0.01 K                   & 13          \\  \hline
Temperature directional coupler & 0.01 K                   & 42      \\
and interferometer arms with & & \\
different transmissions, $\eta=\sqrt{2}$ & & \\  \hline
Frequency of Laser              & 1 GHz                    & 305         \\  \hline
Frequency of Stabilized Laser   & 10 MHz                   & 3.1         \\  \hline
\end{tabular}
	\caption{Influences on the phase difference between the two arms of a fiber interferometer with 1 cm length difference.}
	\label{table1}
\end{table}

\section{Experimental set-up}

\begin{figure}[tb]
\begin{picture}(250,205)
\put(0,0){\psfig{figure=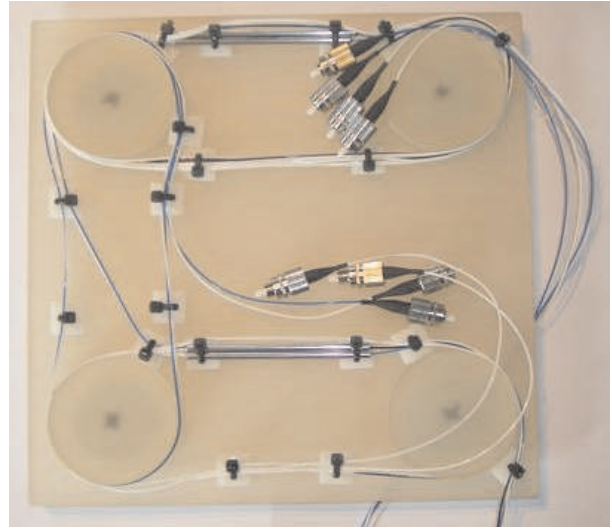,height=70mm,width=80mm}}
\end{picture}
\caption{\label{fig2} Basic set-up of the Mach-Zehnder interferometer. At the bottom and left the {\it effect} interferometer with the perpendicular arms is wrapped around glass discs with diameters of 40~mm. At the top the {\it control} interferometer with the parallel arms is positioned. The four metal bars in the middle of the arms are the directional couplers. The fiber stretchers have been omitted in this photo for clarity reasons.}
\end{figure}

The experiment has been designed with the above raised issues kept in mind. Two interferometers were coupled by means of an additional directional coupler to the same laser (a stabilized He-Ne laser type Coherent 200, linear polarized, 0.5~mW, maximum mode sweep of 10~MHz). One with its arms perpendicular to each other as shown in figure~\ref{fig1}. Another with its arms parallel as a control. Any anisotropy in the speed of light should turn up in the first one and should cancel in the second one. The fibers used were single mode fibers SM600. The five 2x2 directional couplers used were all FC632-50B-FC Split Ratio Coupler 632~nm, 50/50 from Thorlabs Inc. The detectors used are amplified Silicon detectors (PDA36A from Thorlabs Inc). To prevent optical feedback from the interferometer into the laser a polarisation dependent optical isolator (IO-2D-633-VLP from Thorlabs Inc, isolation at least 35~dB) was included (see figure~\ref{fig1}). The length of the fibers in the arms of the first interferometer was 12~m. The effective arm length however was just 4 m as the fibers were mounted on a glass support of 20 by 20~cm and had to be wrapped around glass discs with 40~mm diameter (see also figure~\ref{fig2}). The length of the fibers in the arms of the second interferometer was 2~m, where the effective arm length was 0.7~m. The glass support is made of ultra-low expansion glass (CCZ) with an expansion coefficient of less than $10^{-7}$~1/K. 

\begin{figure}[bt]
\begin{picture}(250,140)
\put(0,0){\psfig{figure=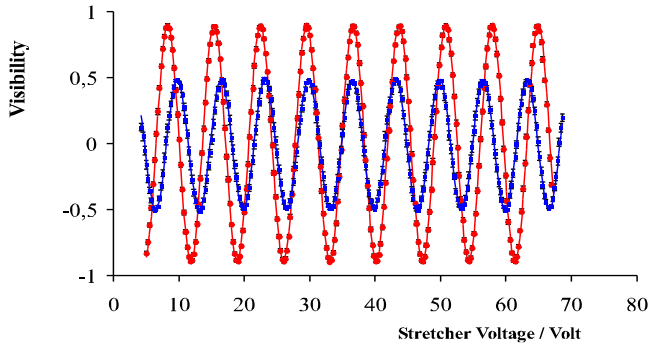,height=45mm,width=85mm}}
\end{picture}
\caption{\label{fig3} Visibility of both interferometers (red {\it control} and blue {\it effect}) as function of the voltage applied to the fiber stretchers.}
\end{figure}

In all arms an extra 1 meter long fiber was inserted wrapped around a thin-walled cylinder made of piezo material acting as a fiber stretcher. For one interferometer the fiber stretchers in the different arms were used with opposite voltage to simultaneously increase the length of the fiber in one arm and reduce it in the other, so that the effect on the phase difference is twice as large. The typical response of the fiber stretchers as function of applied voltage is shown in figure~\ref{fig3}. The lines are cosine fits to the measured points with a frequency of 0.887(1)~rad/V and an amplitude of 0.891(1) for the stretcher in the {\it control} interferometer and 0.940(1)~rad/V repectively 0.492(1) for the other one. The frequency corresponds to a change in the circumference of the cylinders of about 100~nm/V. The standard deviation in the measured visibility is almost as large as the symbols. Note that the visibility amplitude of both interferometers is quite different. This can be due to a bad transmission of one of the arms or a misaligned polarisation~\cite{BookFOS}.

As was shown in the previous section a stable temperature is of the utmost importance. First to ensure that there is no phase change due to a change in temperature and second a stable temperature reduces the possible temperature gradients. Therefor the whole set-up has been positioned in a temperature controlled environment consisting of three compartments. The first compartment is the laboratory, where the temperature is controlled to be constant within $\pm$2~K. The second compartment consists of a box containing the detectors and electronics for the temperature sensors. The walls of this box are made of 1~mm thick aluminum shields, 5~cm of a poly urethane foam isolation and again an aluminum shield of at least 1~mm thickness. The heat conductivity of aluminum is 150~W/(m K) and of poly urethane 0.038~W/(m K). This large difference ensures a homogeneous distribution of the temperature. In this box the air temperature is controlled to be constant within $\pm$0.2~K. The third compartment consists of a smaller box containing the interferometers and optical isolator. The walls are constructed in a similar way as the walls of the second compartment for the same reason. The temperature of this box is stable for days within 3~mK. For shorter times the temperature is stable within 1~mK. The heaters are glued to the outside of the aluminum shields. The temperature sensors are PT100s. The heat applied to control the temperature of the middle compartment varies between 1 and 4~W. The heat applied to control the temperature of the inner compartment is below 150~mW. This ensures that the possible temperature gradients within the inner compartment are less than 1~mK/m.

The second compartment together with the laser and control electronics are put on a rotation stage (LT360 Precision Turntable from LinearX Systems Inc with a rotation accuracy of 0.1~degree) to enable the change of azimuth. 

\begin{figure}[tb]
\begin{picture}(250,390)
\put(8,260){\psfig{figure=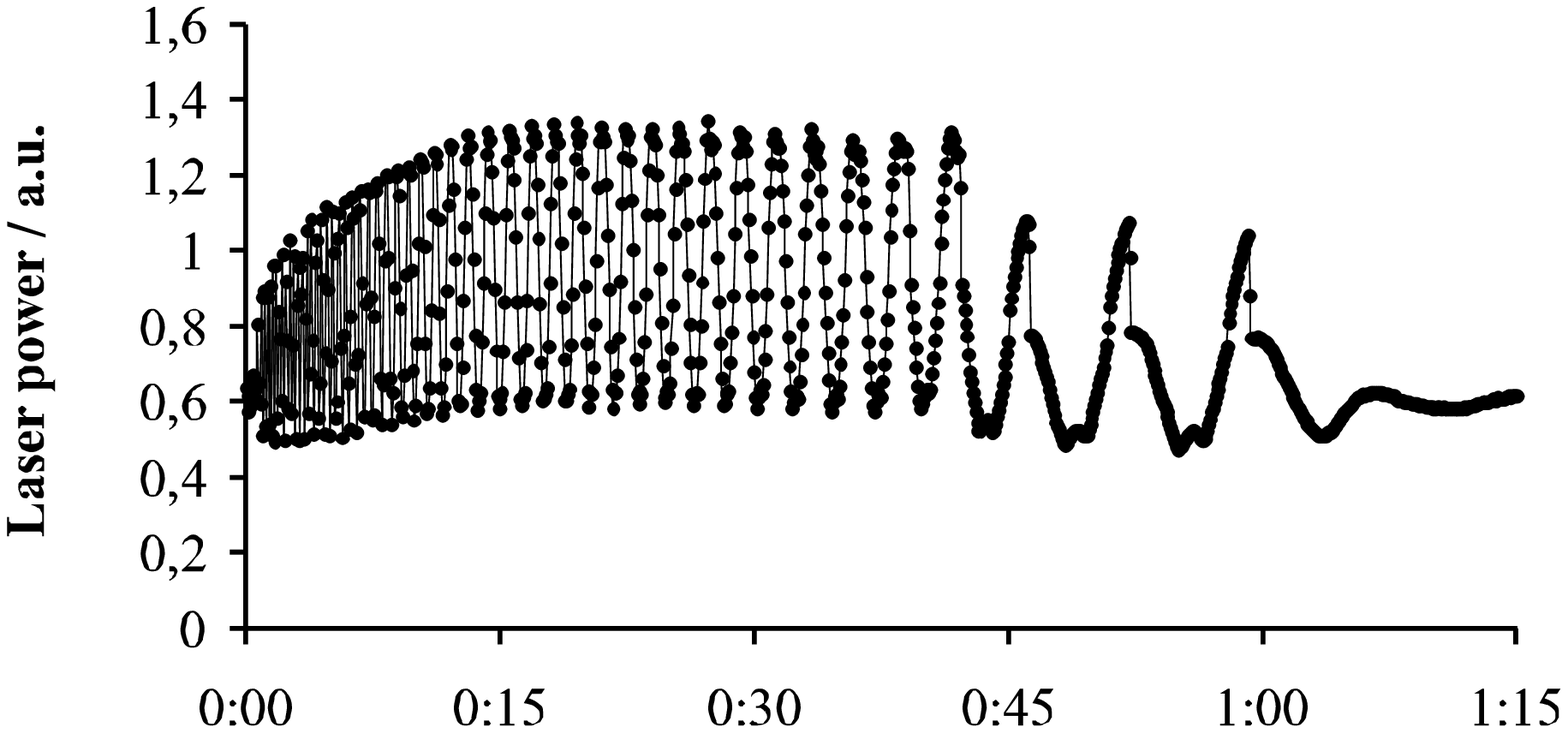,height=45mm,width=80mm}}
\put(0,130){\psfig{figure=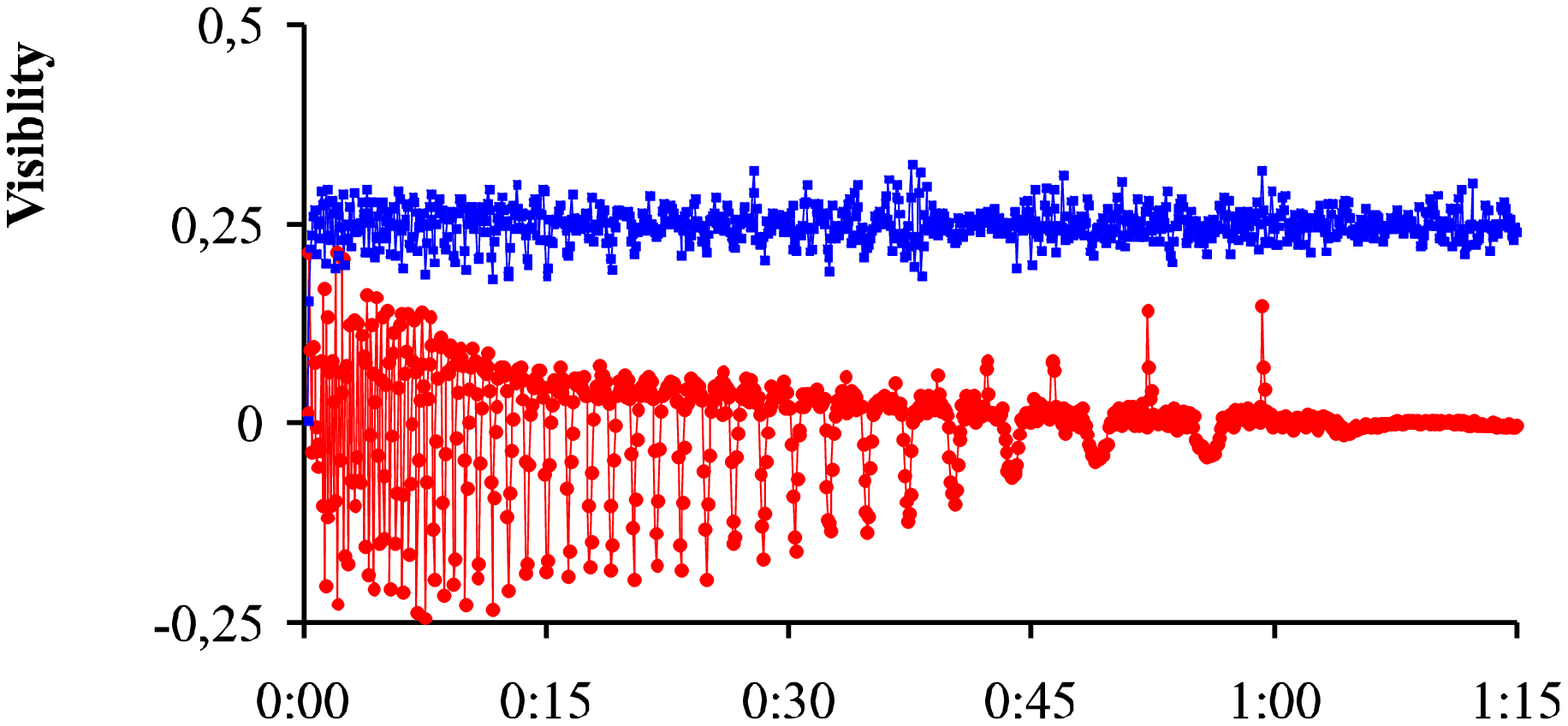,height=45mm,width=80mm}}
\put(4,2){\psfig{figure=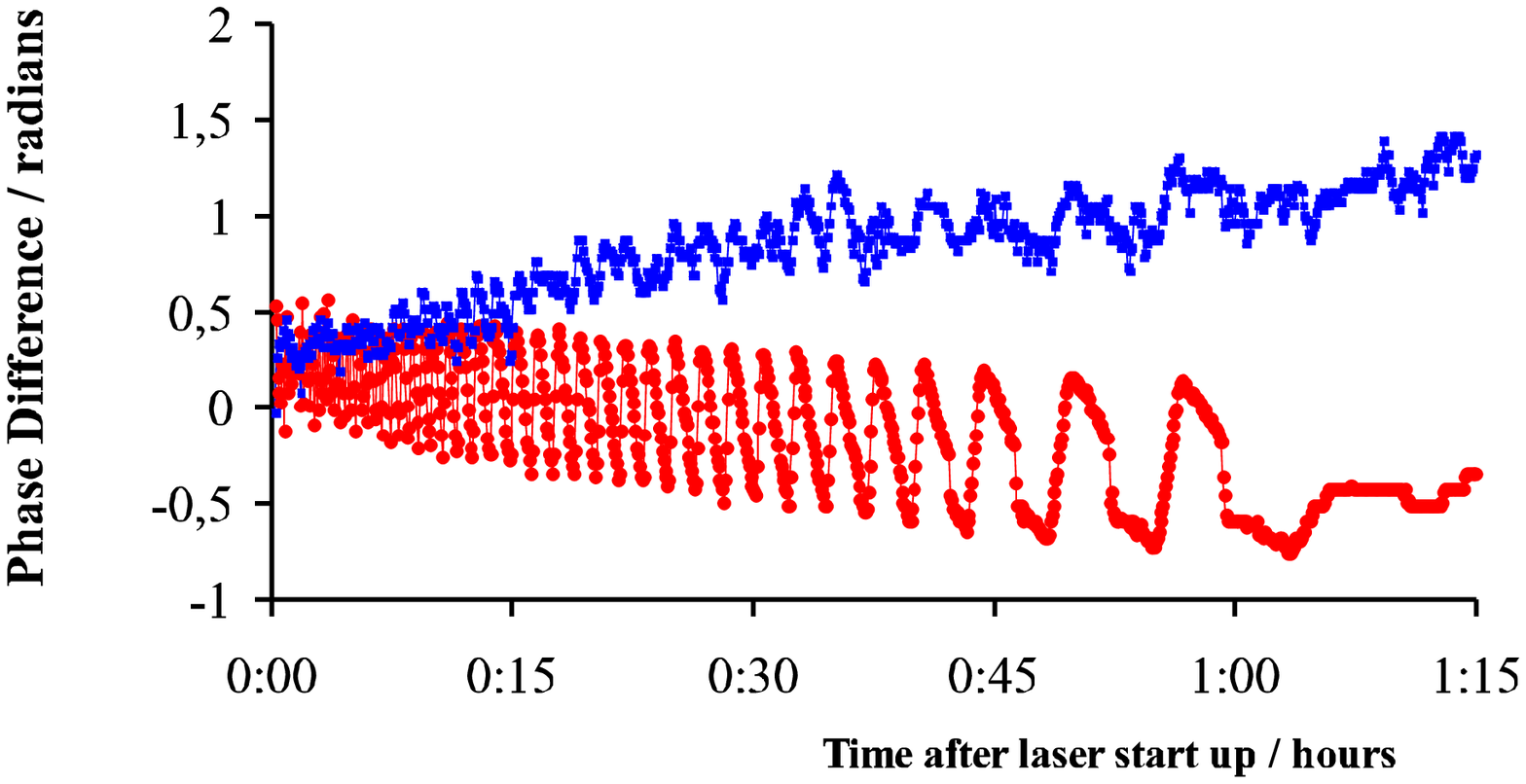,height=45mm,width=80mm}}
\end{picture}
\caption{\label{fig3B} Top: laser power as function of warm up time showing mode changes as changing intensities. Middle: Visibility of both interferometers (red {\it control} and blue {\it effect}, shifted with +0.25) during the same period. Bottom: Phase difference of both interferometers during the same period, showing the influence of the changes of 0.565~GHz in the frequency of the laser.}
\end{figure}

\section{Results and discussion}

First, to find the effect of the laser frequency on the phase difference of the interferometers, the laser was switched off and switched on again after a cool down period of one hour. It takes the laser sometime to become stabilized again. During this warm up period the laser experiences regular mode changes, depending on the size of laser cavity. These mode changes of 0.565~GHz change the phase difference of the interferometers and are shown in figure~\ref{fig3B}. The laser was turned on and after 2 minutes the phase control was turned on. The visibilities then rapidly decrease to around 0 and the phase difference is controlled by the stretchers. The visibility then becomes an error signal, indicating the accuracy of the controlled phase. Due to a time constant of the control of 1 s, rapid changes in phase difference create an error signal, which is used to change the voltage applied to the shifters. From this graph it is clear that both interferometers are sensitive to a change in laser frequency. For the {\it control} interferometer it is 0.86 radians or 1.5 rad/GHz, for the other its about half of that value. Hence, their sensitivity to a frequency change is much larger than expected due to a difference in length of the arms of the interferometers.  The reason for this discrepancy is sought for in the interaction length of the directional couplers. However, as long a the laser is stabilized the mode changes are less than 10~MHz, reducing the effect to negliable proportions.

\begin{figure}[tb]
\begin{picture}(250,260)
\put(0,130){\psfig{figure=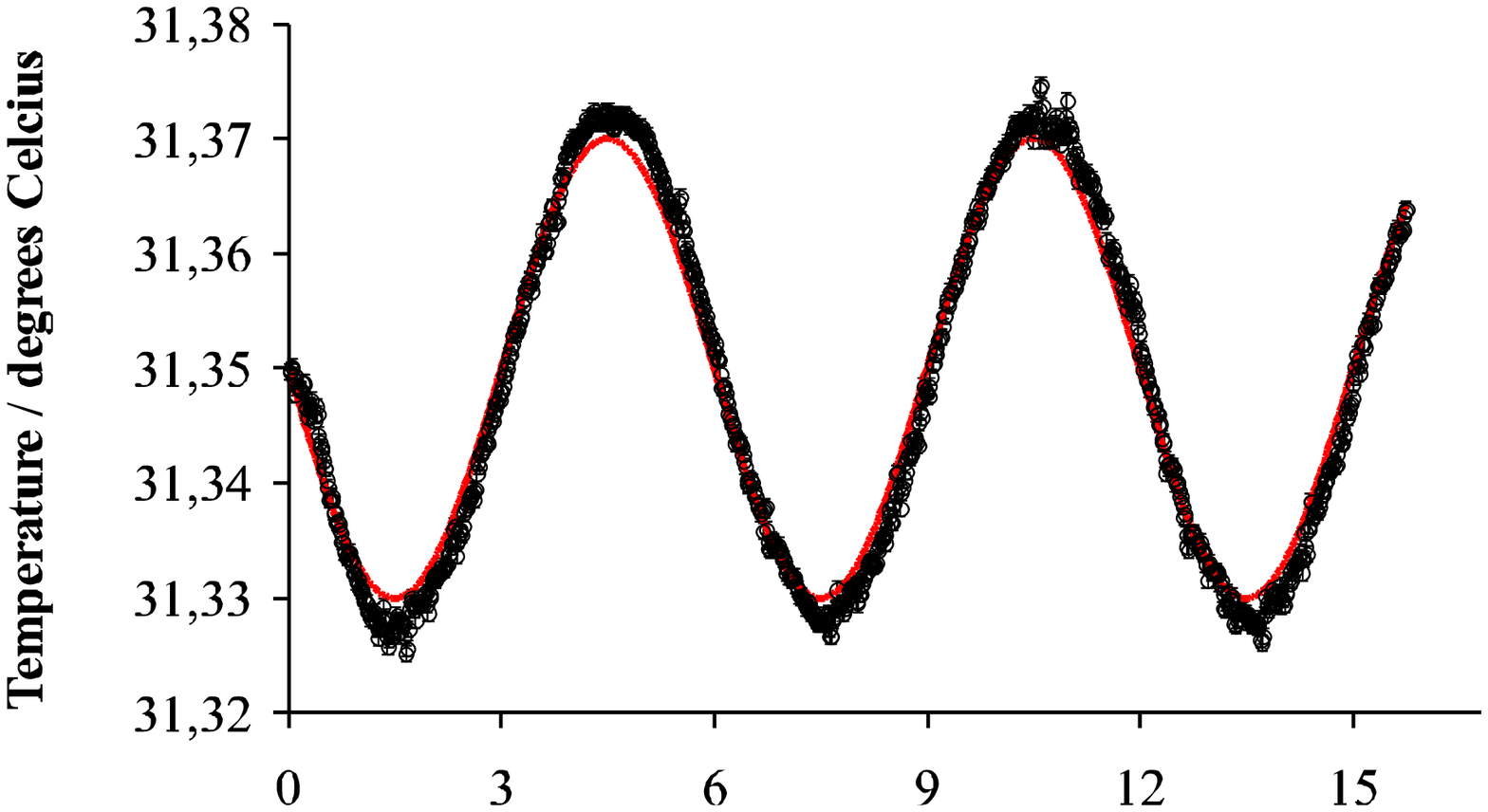,height=45mm,width=80mm}}
\put(4,2){\psfig{figure=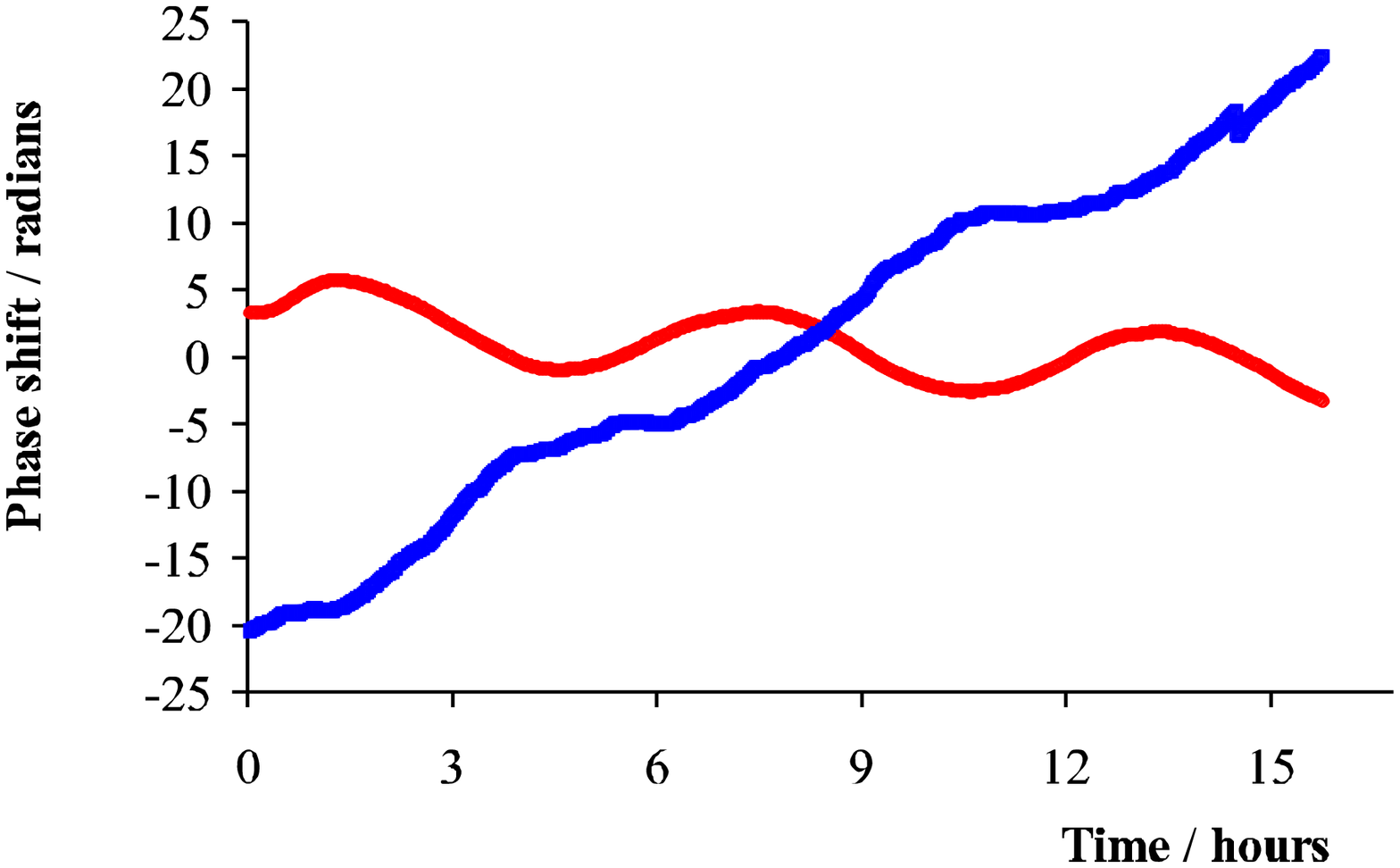,height=45mm,width=80mm}}
\end{picture}
\caption{\label{fig4} Top: set-point (red line) and control temperature as function of time. Bottom: phase difference of interferometers (red {\it control} and blue {\it effect}) during the same period as the upper graph. The phase oscillations due to the temperature oscillations are reproduced imposed on a linear drift of unknown origin.}
\end{figure}

Second, the influence of the temperature on the phase difference has been determined by varying the temperature of the most inner container as a cosine with an amplitude of 0.02 K and a period of 6 hours. The results are shown in figure~\ref{fig4}. 
The red line in the top graph shows the set-point temperature. The black circles and error bars represent the control temperature. The deviations are due to the limitations of the temperature control. The curves in the bottom graph show the corresponding oscillations of the phase difference of the {\it control} interferometer (in red) and for the other one (in blue). The oscillations correspond to a sensitivity to temperature variations of -130~rad/K of the {\it control} interferometer and for the other one to a sensitivity of +90~rad/K. These values are much large than estimated in section~\ref{secinf}. This could indicate that the coupling length of the directional couplers is much larger than assumed, or that there remains a considerable temperature gradient over the arms that changes for a changing temperature. The fact that such a small and slow change of temperature of the interferometer results in such a large change in the phase difference, much larger than expected from the simple considerations above, could explain the results of Cahill~\cite{Cahill2008} as he took great care to minimize the temperature gradients, but did not control the temperature of his instrument.
An extra linear decrease or increase in the phases as function of elapsed time is also observed. For the {\it effect} interferometer, this change is larger than for the {\it control} interferometer. This is typical for all experiments done with these interferometers, indicating that building a fiber interferometer with a stability over days is very hard to accomplish. The mechanisms causing these changes have not been explicitly determined. A possibility is the long time behavior of fibers properties and fiber stretcher properties. For instance, the fibers used are glass fibers and glass essentially remains a very thick liquid. Further, the stretchers are made of piezo material that has some long term relaxation. 

\begin{figure}[tb]
\begin{picture}(250,350)
\put(2,0){\psfig{figure=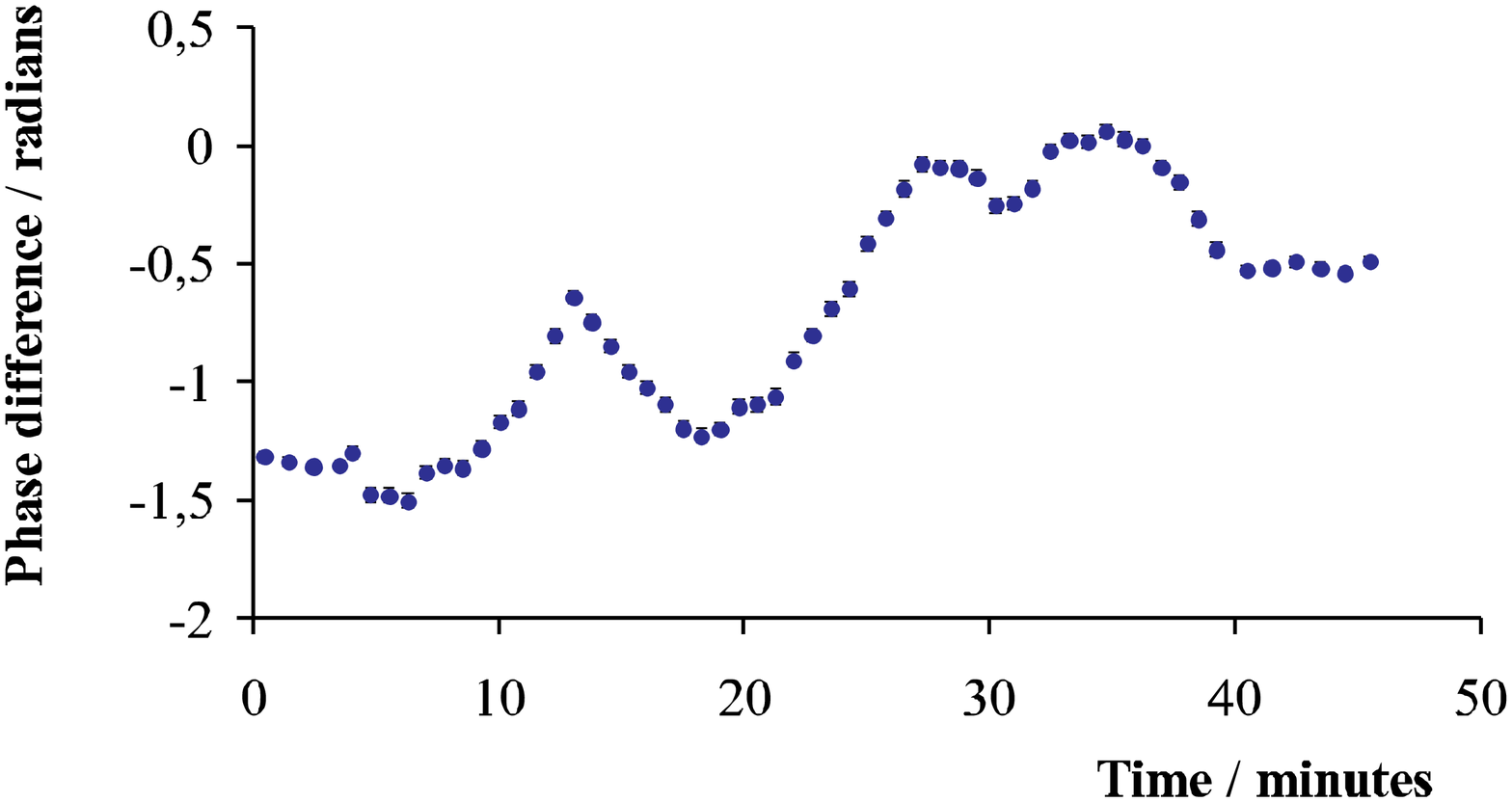,height=40mm,width=80mm}}
\put(0,115){\psfig{figure=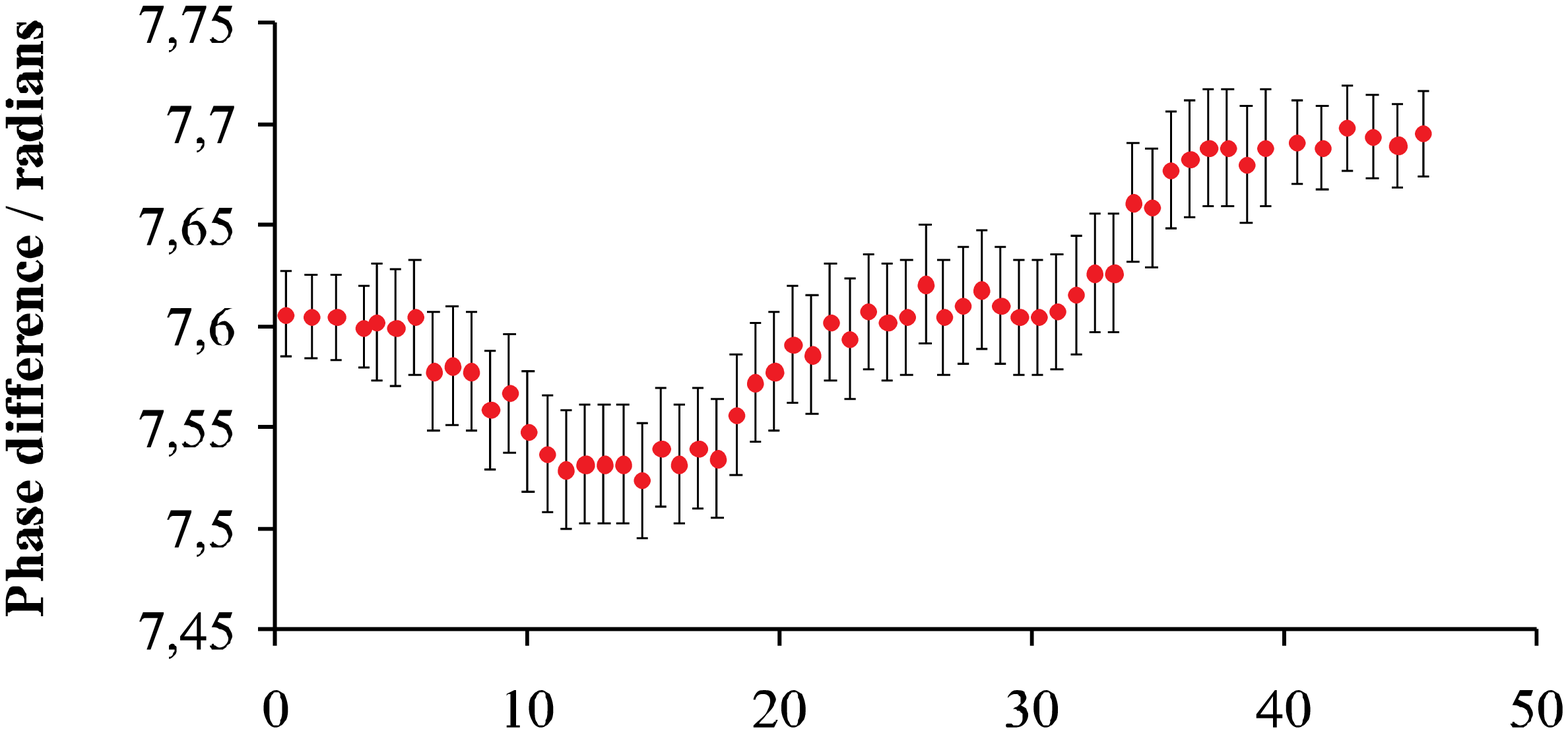,height=40mm,width=80mm}}
\put(3,230){\psfig{figure=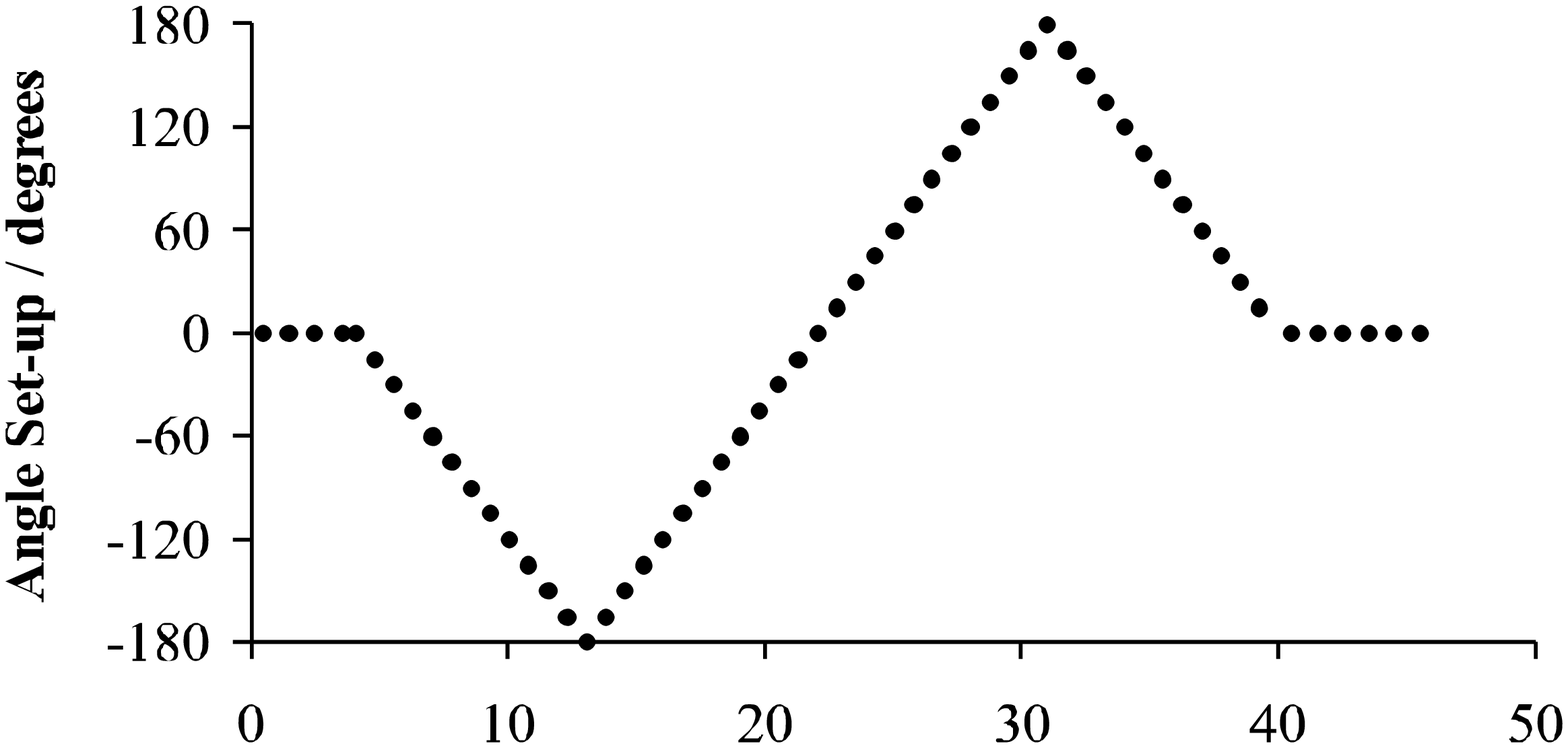,height=40mm,width=80mm}}
\end{picture}
\caption{\label{fig5} Top: angle of set-up as function of time. Middle: phase difference of {\it control} interferometer during the same period as the upper graph. Bottom: phase difference of the {\it effect} interferometer again during the same period.}
\end{figure}

\begin{figure}[tb]
\begin{picture}(250,240)
\put(0,120){\psfig{figure=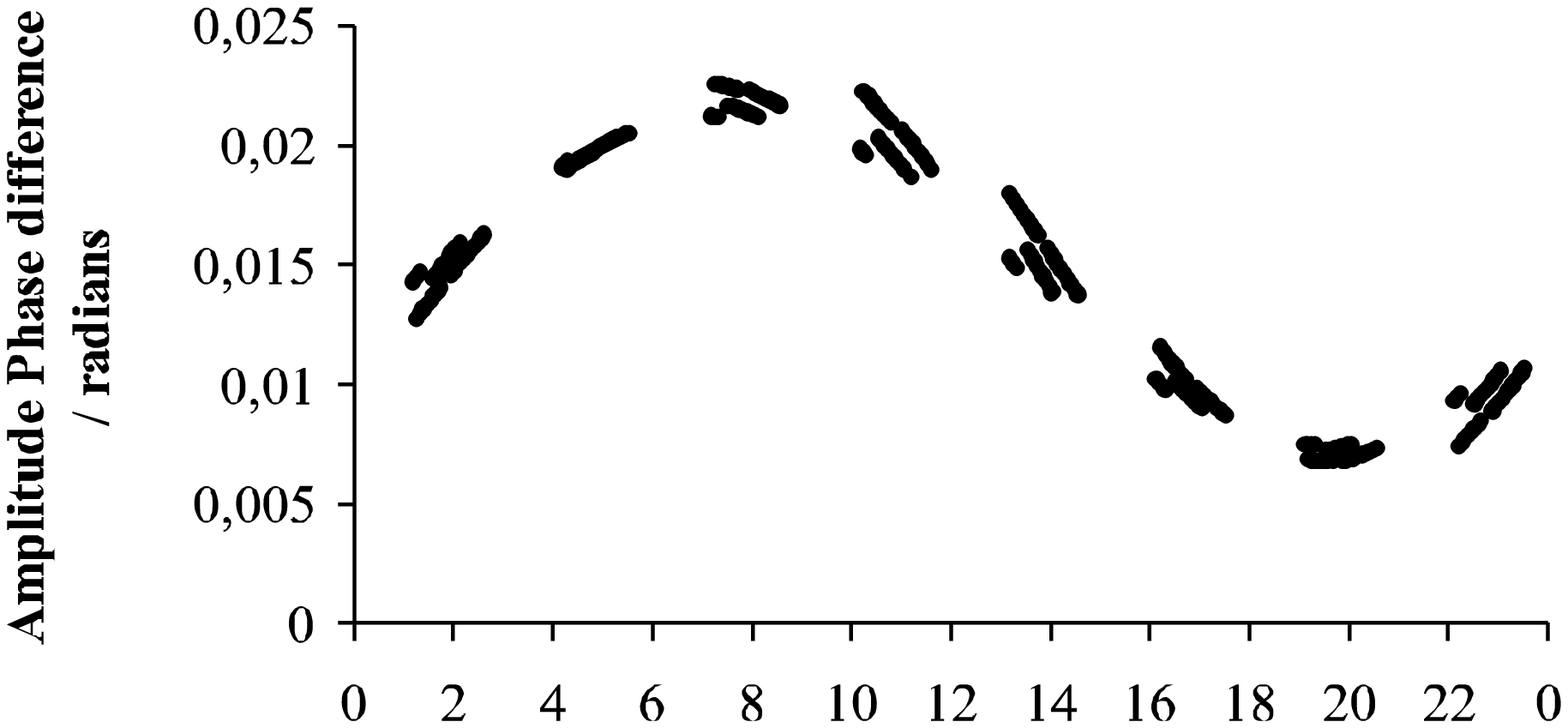,height=40mm,width=80mm}}
\put(7,0){\psfig{figure=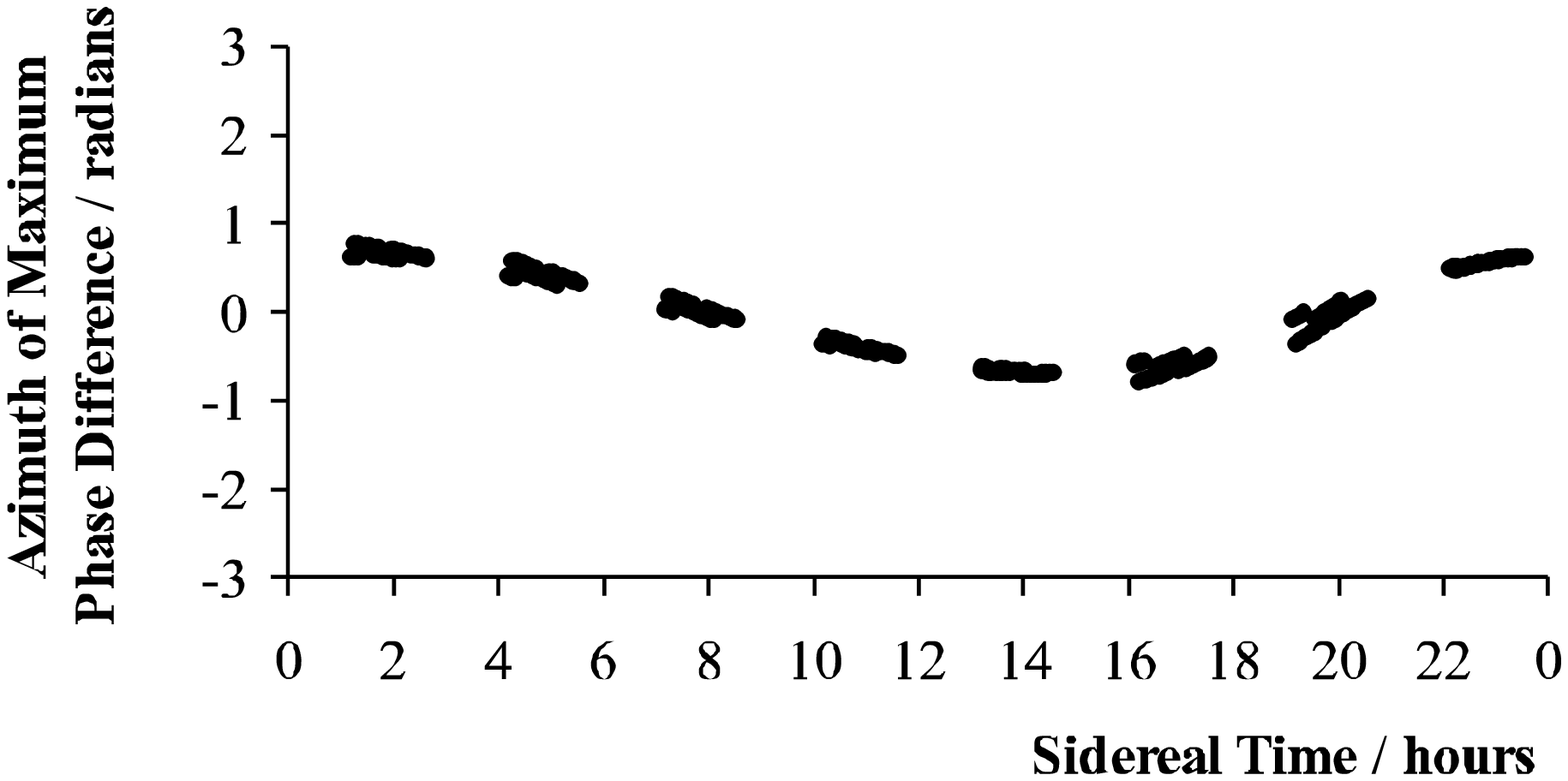,height=40mm,width=80mm}}
\end{picture}
\caption{\label{fig6ab} Top: second order amplitude of the signal of the {\it effect} interferometer as function of sidereal time as calculated for the same times and epoch as with the experiment. Bottom: second order azimuth of the maximum of the same signal.}
\end{figure}

Finally, the set-up is rotated along a vertical axis to find out if there is an influence due to the velocity of the Earth in a preferred rest frame. Every 3 hours the set-up rotates from 0 to -180 degrees, from -180 to +180 and from +180 to 0 with steps of 15 degrees. For 0 degrees the interferometer with the parallel arms points to the local North. For 90 degrees it points to the local East. The longitude and latitude of the location of the interferometers on Earth are 4.7 degrees and 51.4 degrees. At every step the phase of both interferometers is measured during half a minute to average out statistical variations. A complete rotation takes about 40 minutes. A typical response of the interferometers as function of angle is shown in figure~\ref{fig5}.
The top graph of this figure shows the angle of the set-up as function of time. The middle one shows the phase difference of the {\it control} interferometer during the same period as the upper graph. The bottom graph shows the phase difference of the {\it effect} interferometer again during the same period. In the lower graph the phase oscillations due to the rotation of the set-up are clearly visible, again imposed on a linear drift of unknown origin. At first, it seems that this interferometer gives an indication that there might be some effect on the phase difference due to the motion of the Earth. However, if such an effect exists it should depend on the time of day and year. Upon rotation of the set up, the amplitude and azimuth of the maximum should vary between certain minima and maxima depending on the orientation of the Earth velocity with respect to the preferred frame as discussed by M\'{u}neara~\cite{Munera1998}. The Fourier transform of the phase difference (corrected for the linear assumed drift) as function of rotation angle gives this amplitude and azimuth of the maximum phase difference for all orders. The zeroth order is just the average phase difference during a rotation. The first order represents the amplitude and azimuth of that part of the signal that varies with the cosine of the angle of the set-up, corresponding to first order effects in $v/c$. The second order represents the amplitude and azimuth of that part of the signal that varies with the cosine of twice the angle of the set-up, corresponding to second order effects, and so on. The error in the values can be estimated from the difference between the Fourier transform of the data points measured for increasing set-up angles and the one measured for decreasing set-up angles (to find the systematic error due to the unknown drift) combined with the Fourier transform of the variances (to find an estimate of the statistical error). The amplitude and azimuth should vary with the sidereal time and epoch. For the month March 2009 the effect is calculated using the same rotations of the set-up as used in the experiment. The theoretical values were treated in exactly the same way as the measurement data were treated, except that instead of using the measured results for the phase difference, the theoretical values for the phase differences are used. The theoretical values were calculated following M\'{u}nera~\cite{Munera1998} using the Miller data for the velocity of the Sun (Right ascension 4.9~hours and declination -70.6~degrees with a velocity of 205~km/s). Miller assumes that due to some unknown cause the effect is reduced by a factor of 20~\cite{Miller1933} in the velocity. Here the effect is reduced by a factor of 100 and hence the second order term coefficient becomes $0.0001\times 4\pi nL/\lambda_0 = 1.2\times 10^4 $. The results are shown in figure~\ref{fig6ab}. A clear dependence of the amplitude and maximum azimuth appears.
The irregular shape of the curves is due to the change of the projection of Earth velocity with respect to the preferred frame on surface of the Earth during one rotation (as a complete rotation of the set-up takes about 40 min). During the experiment the set-up was rotated every 3 hours. This would cover 2/3 of the sidereal time scale in one month as the difference between a sidereal day and a normal day is 1/365 of a day, which adds up to 2 hours per month. However, at the middle of the series, the rotation interval was started again, shifting the times at which the rotations occured so that an overlap occured with previously measured sidereal time. The same happened again at the end of the series when the computer clock was advanced with one hour due to the end of daylight saving time. The results is a smaller coverage of the sidereal time scale and not overlapping results. The slightly different values for sidereal times that are the same, are due to the difference of the position of the Earth in its orbit around the Sun at the time the data point were taken. Hence, for instance, the effect for 10 hours sidereal time at the beginning of the month is a bit different from the effect for 10 hours sidereal time at the end of the month. During a year this leads to much larger differences~\cite{Munera1998}, and should be taken into account for the analysis of these kinds of measurements. However, during the observation period, these irregularities are quite small with respect to the full oscillation of the effect during a sidereal day and are further ignored. 

\begin{figure}[tb]
\begin{picture}(250,240)
\put(0,120){\psfig{figure=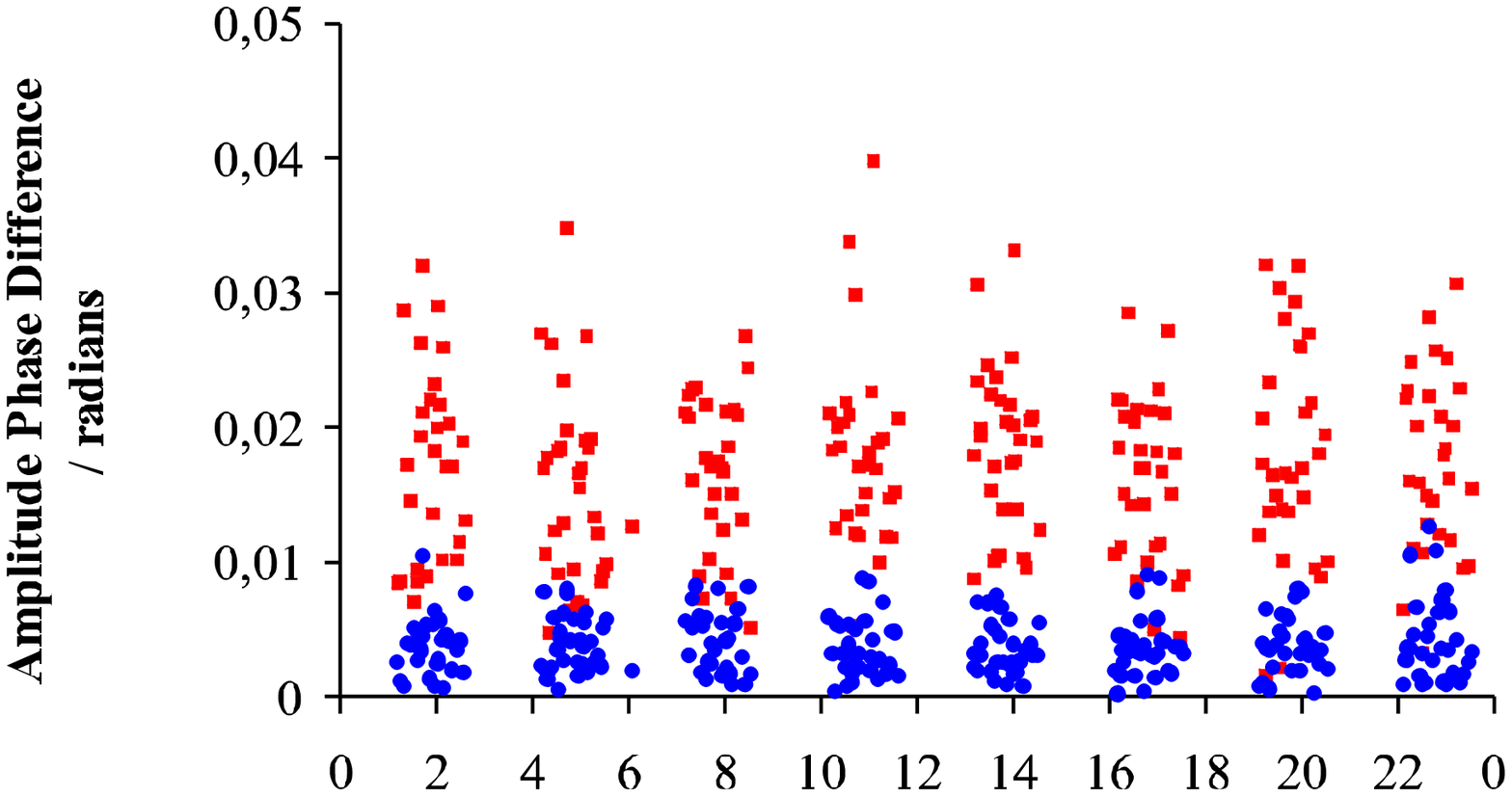,height=40mm,width=80mm}}
\put(7,0){\psfig{figure=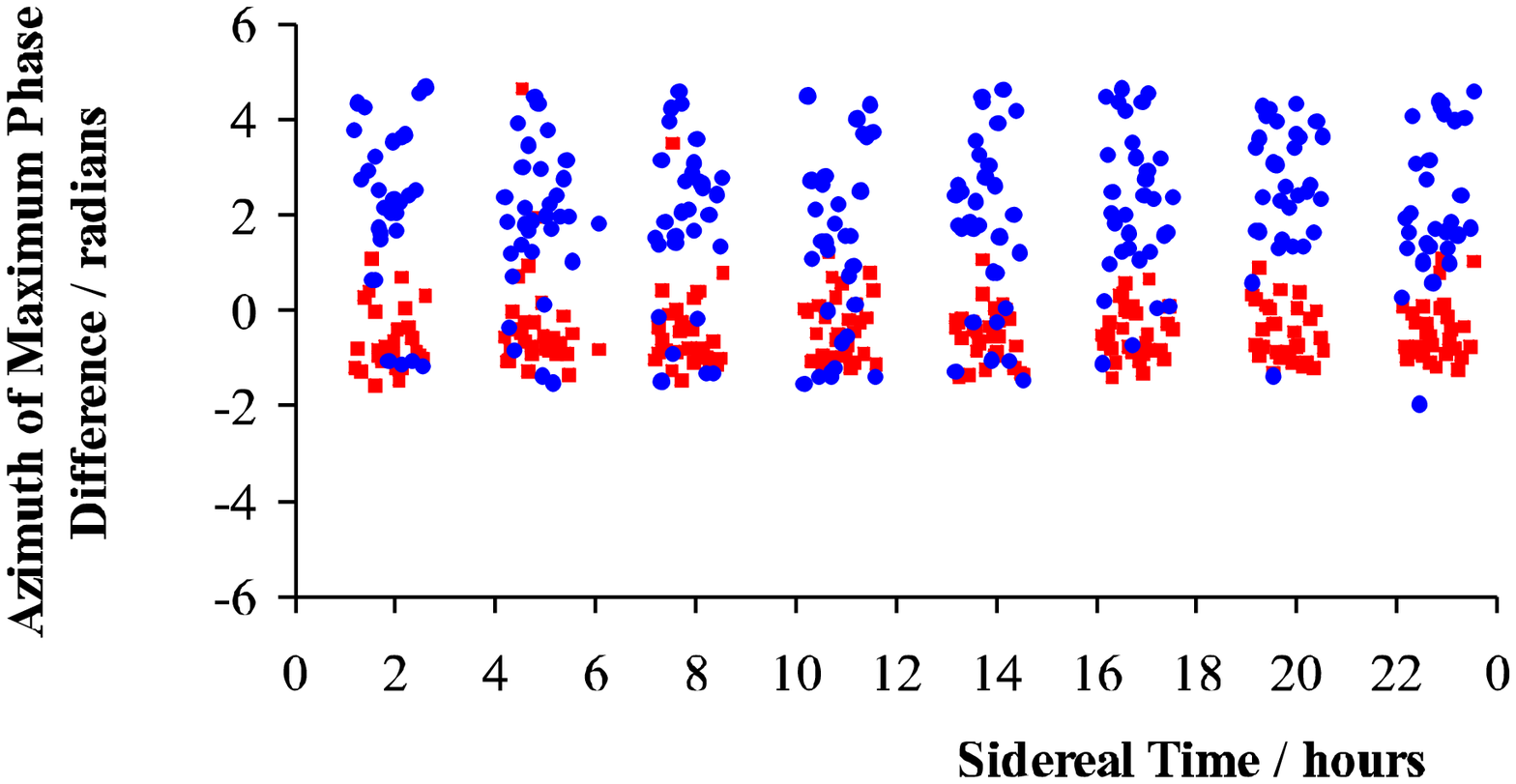,height=40mm,width=80mm}}
\end{picture}
\caption{\label{fig7ab} Top: first (red) and second (blue) order amplitude of the signal of the {\it control} interferometer as function of sidereal time. Bottom: first (red) and second (blue) order azimuth of the maximum of the same signal. For clarity reason the error bars have been omitted. They have approximately the same length as the spread in points.}
\end{figure}

\begin{figure}[bt]
\begin{picture}(250,240)
\put(0,120){\psfig{figure=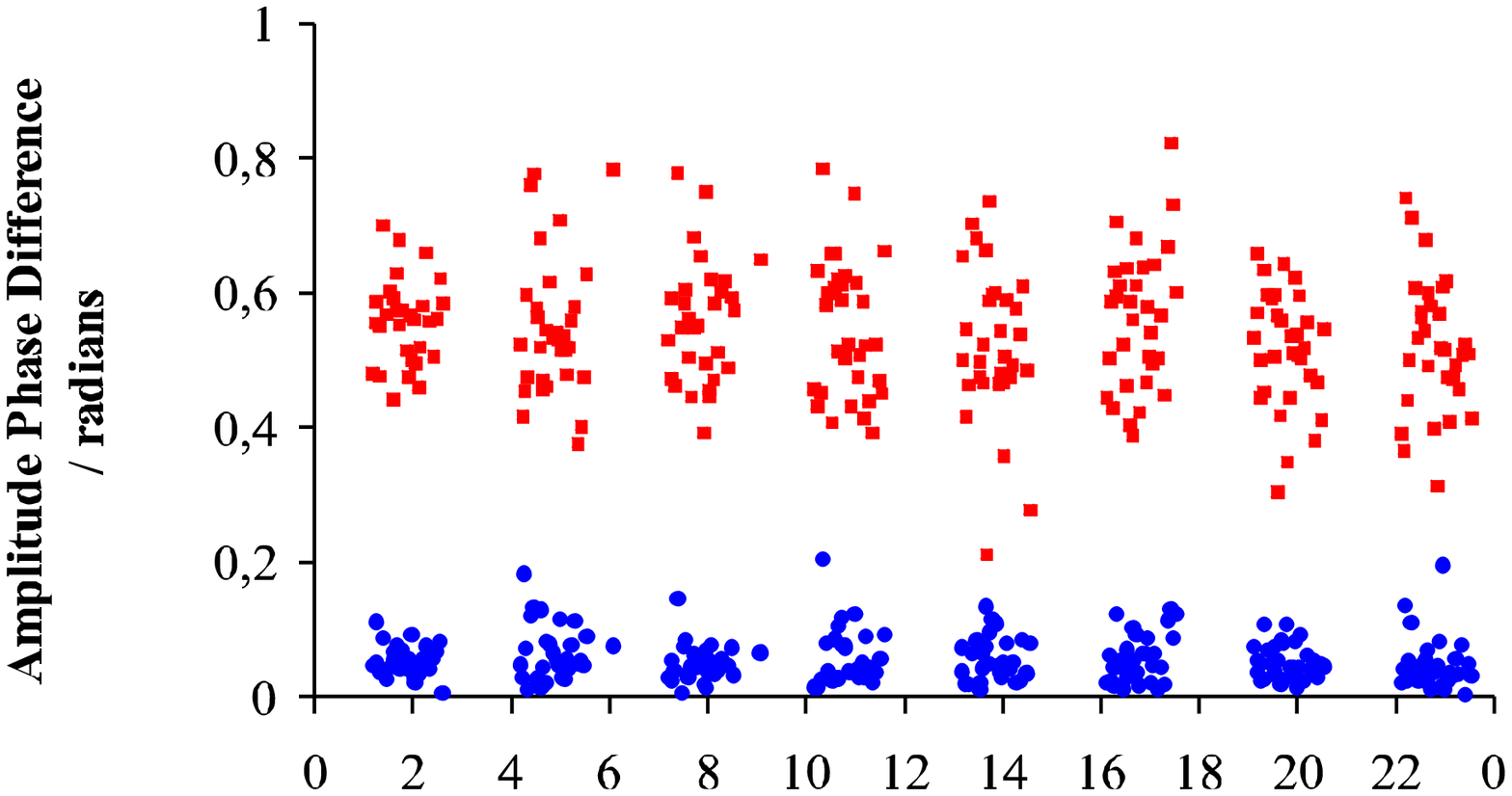,height=40mm,width=80mm}}
\put(4,0){\psfig{figure=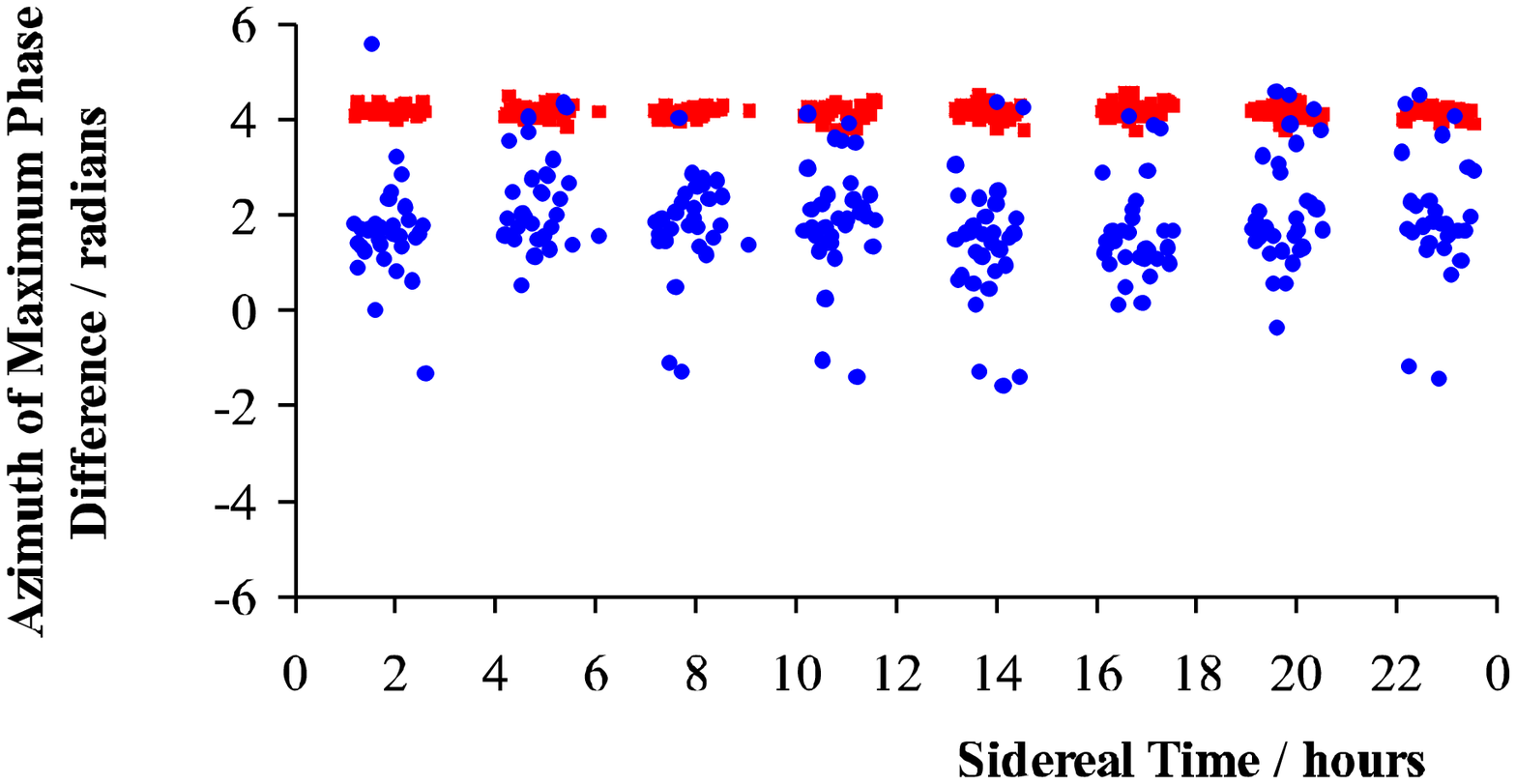,height=40mm,width=80mm}}
\end{picture}
\caption{\label{fig8ab} Top: first (red) and second (blue) order amplitude of the signal of the {\it effect} interferometer as function of sidereal time. Bottom: first (red) and second (blue) order azimuth of the maximum of the same signal. For clarity reason the error bars have been omitted. They have approximately the same length as the spread in points.}
\end{figure}

For the {\it control} interferometer the amplitude and azimuth of the maximum measured as function of sidereal time in the month March 2009 are shown in figure~\ref{fig7ab}. Similar data for the {\it effect} interferometer are shown in figure~\ref{fig8ab}. There is no apparent sidereal signal in any of the graphs. Hence, the conclusion is justified that the built Mach-Zehnder fiber interferometer does not measure any anisotropy of the velocity of light on the Earth surface.

When the second order signal of the interferometer with the long fibers is interpreted in a similar way as was done for the Michelson-Morley experiments, the result would be that the second order amplitude of the signal is less than 0.06 radians corresponding to 0.01 parts of a fringe. Using the simple formula $\Delta \phi=4\pi nL/\lambda_0(v/c)^2$ the result is a maximum possible velocity of 7~km/s. This is comparable to the results obtained from all previous Michelson-Morley type interferometer experiments.     

\section{Conclusions}

The two optical fiber interferometers built in a temperature controlled environment enable the determination of temperature effects on the phase difference between the light beams traveling through the two arms of the interferometer. The temperature influence on this phase difference is much larger than theoretically expected. It has been shown, that altough at first sight the temperature has no or limited influence on the working of a directional coupler, in combination with a variation in another parameter, in this case the transmission difference of the arms, it can have an important influence. So the measured phase difference could be due to the interaction region of the directional coupler or to a temperature gradient of unknown origin. 

Upon rotation of the interferometers around a vertical axis in one of the interferometers an oscillation in the phase difference as function of the azimuth of the set-up is observed. With regard to the measurement accuracy, this oscillation is constant and is attributed to changing stresses in the arms of the interferometer.
Analysis of the signal of the interferometer with the perpendicular arms show that the fiber optical Mach-Zehnder interferometer built, is unable to detect an anisotropy of the velocity of light at the Earth surface. When the signal is interpreted in a similar way as Michelson and Morley did in their famous experiment, the resulting maximum velocity of Earth with respect to the preferred frame would be 7~km/s.

In view of the experimental difficulties getting a stable signal it is questionable that the accuracy of this type of measurements of the anisotropy of the speed of light on the Earth surface could be increased.

\end{document}